\newcommand{\be}{\begin{eqnarray}}
\newcommand{\ee}{\end{eqnarray}}

\newcommand{\bea}{\begin{eqnarray}}
\newcommand{\eea}{\end{eqnarray}}
\newcommand{\beas}{\begin{eqnarray*}}
\newcommand{\eeas}{\end{eqnarray*}}
\documentclass[twocolumn,aps,showpacs,preprintnumbers,nofootinbib]{revtex4}
\usepackage{color}
\usepackage{graphics}
\usepackage{epsfig}
\usepackage{amsmath} 
\usepackage{amssymb}
\usepackage{amsthm}
\begin{document}
\title{Three and two-hadron correlations in $\sqrt{s_{NN}}=200$ GeV
  proton-proton and nucleus-nucleus collisions} 

\preprint{BCCUNY-HEP/09-09}

\author{Alejandro Ayala$^{1,2}$, Jamal Jalilian-Marian$^3$, J. Magnin$^2$,
  Antonio Ort\1z$^1$, G. Pai\'c$^1$ and Maria Elena Tejeda-Yeomans$^4$}   
\affiliation{$^1$Instituto de Ciencias Nucleares, Universidad
Nacional Aut\'onoma de M\'exico, Apartado Postal 70-543, M\'exico
Distrito Federal 04510, M\'exico.\\
$^2$Centro Brasileiro de Pesquisas F\1sicas, CBPF,
Rua Dr. Xavier Sigaud 150, 22290-180, Rio de Janeiro, Brazil.\\
$^3³$Department of Natural Sciences, Baruch College, New York, NY 10010, USA\\
and CUNY Graduate Center, 365 Fifth Ave., New York, NY 10016, USA.\\
$^4$Departamento de F\1sica, Universidad de Sonora, Boulevard Luis Encinas
J. y Rosales, Colonia Centro, Hermosillo, Sonora 83000, M\'exico.}

\begin{abstract} 

We compare the azimuthal correlations arising from three and two hadron
production in high energy proton-proton and nucleus-nucleus collisions at
$\sqrt{s_{NN}}=200$ GeV, using the leading order matrix elements for
two-to-three and two-to-two parton-processes in perturbative QCD. We first
compute the two and three hadron production cross sections in mid-rapidity
proton-proton collisions. Then we consider Au + Au collisions including parton
energy loss using the modified fragmentation function approach. By examining
the geometrical paths the hard partons follow through the medium, we show that
the two away-side partons produced in two-to-three processes have in average a
smaller and a greater path length than the average path length of the
away-side parton in two-to-two processes. Therefore there is a large
probability that in the former processes one of the particles escapes while
the other gets absorbed. This effect leads to an enhancement in the azimuthal
correlations of the two-to-three with respect to the two-to-two
parton-processes when comparing to the same processes in proton-proton
collisions since in average the particle with the shortest path length looses
less energy with respect to the away side particle in two-to-two processes. We
argue  that this phenomenon may be responsible for the shape of the away-side
in azimuthal correlations observed in mid-rapidity Au + Au collisions at
RHIC.

\end{abstract}

\pacs{25.75.-q, 25.75.Gz}

\maketitle

\date{\today}

Many interesting phenomena have been observed in high energy heavy ion
collisions at the Relativistic Heavy-Ion Collider (RHIC) since it began
operation almost a decade ago. The suppression of the single hadron transverse
momentum spectra in Au + Au collisions, as compared to normalized p + p
collisions is one of such phenomena. This is believed to result from
the energy loss of fast partons traversing the medium and multiply scattering
(both elasticly and inelasticly) from medium constituents~\cite{eloss, urs}.
This picture of energy loss 
was further confirmed by azimuthal correlation studies where for certain
combinations of leading and associated particles a disappearance of the 
away-side peak is observed. This is interpreted as the absorption of the 
away side parton by the medium. Combined with an enhancement of the single 
hadron production in d + Au collisions as compared to normalized p + p
collisions, 
these observations established the presence of a final state medium produced
in Au + Au collisions at RHIC. Furthermore, the strong collectivity of the
medium, as measured by the flow parameter $v_2$, among other 
observations, have led to the consensus that the medium is
partonic in origin as well as strongly interacting~\cite{rhic_exp, gm}. 

In spite of the success of the above interpretation of RHIC data~\cite{onehad}, 
there still remain important aspects of energy loss dynamics which are poorly
understood. 
The picture has been further complicated by the recent observation in
azimuthal correlation studies in  Au + Au collisions, where for the case when
the magnitude of the momentum difference between leading and associated
particles increases, either a {\it double hump} structure or a broadening of
the away-side peak appears, while these structures are absent in p + p
collisions at the same energy~\cite{PHENIX, STAR}. This observation gave rise
to a large number of theoretical explanations which are based on the 
assumption that collective phenomena are at work in A + A collisions in
contrast to the case of p + p collisions~\cite{Casalderrey, miklos, Majmunder,
Chiu, Chaudhuri, Takahashi}. Most notably, in Ref.~\cite{Takahashi} a
hydrodynamic approach is used to attribute the double hump structure in the
away side jet to the propagation of pressure gradients from dense zones in the
plasma which can only be observed by averaging over a large number of random
initial conditions and after subtracting the flow component of the azimuthal
correlation function.

However, one can ask whether these features are already present at the level
of p + p collisions, albeit obscured by the smallness of their
intensity. For instance, two-to-three parton processes 
would lead to such a double hump structure even in p + p collisions,
although these would be suppressed with respect to two-to-two
parton processes.
The question is whether in A + A collisions there exists a mechanism that
amplifies hadron production from two-to-three processes with respect to that
from two-to-two parton processes. This 
question has been recently partially investigated in Ref.~\cite{Ayala} 
using the event shape analysis in p + p collisions to distinguish events
containing three jets from those containing two. 

In this Letter, we study angular correlations between three hadrons produced
both in p + p and A + A collisions at RHIC using the full Leading Order (LO)
matrix elements~\cite{lo} for two-to-three parton processes and compare these
to angular correlations between two hadrons using also LO matrix elements for
two-to-two parton processes. To the best of 
our knowledge, this is the first study of its  kind using the complete LO
matrix elements to calculate three hadron production in A + A collisions. 
We should mention that the implementation of Next-to-Leading Order (NLO)
matrix elements for two-to-three processes requires the corresponding nuclear
parton distributions and medium modified fragmentation functions to matching
accuracy. The former are available in the literature~\cite{lo},
the latter are not fully known. Therefore here we use the LO matrix elements,
given that we focus on the nuclear effects. We use a medium  
modified fragmentation function in order to treat the medium-induced parton
energy loss, following the approach of Ref.~\cite{zoww} where two-hadron
production cross section in A + A collisions is studied. We argue that since
in A + A collisions, the final state partons with the short (long) trajectory
in the away side has a large probability to punch through (get absorbed), the
effect produces, on the average, a double hump in the azimuthal correlation. 


Using the LO matrix elements for two-to-three parton processes, including the
final state phase space factors and enforcing momentum conservation, the three
hadron production cross section can be written as
\begin{widetext}
\be
   \frac{d\sigma^{pp \rightarrow h1\, h2\, h3\, X}}{d\, y_1 d\, y_2 d\, y_3 d\,
   h_{1t}  d\, h_{2t} d\, h_{3t} d\, \phi_2 d\, \phi_3 } &= &\frac{1}{2^5}
   \frac{1}{(2\pi)^4}\, \frac{1}{h_{3t}\, S} \, 
   \frac{|\sin \phi_2|}{[\sin \phi_2 \,
   \sin^2 (\phi_3/2) - \sin \phi_3 \, \sin^2 (\phi_2/2)]^2}  
   \int d z_3  |{\mathcal{M}}|^2 \nonumber \\ 
   &\times& f_{i/p} (x_1, \mu^2 ) 
   f_{j/p} (x_2, \mu^2 ) D_{h1/k}^0 (z_1, \mu^2 )
   D_{h2/m}^0 (z_2, \mu^2 ) D_{h3/n}^0 (z_3, \mu^2 ) ,
\label{eq:pp}
\ee
\end{widetext}
where $y_i, h_{it}$ are the rapidity and magnitude of transverse momentum of
the $i$-th produced hadron, $\phi_2 \,  (\phi_3)$ is the azimuthal angle
between the second (third) hadron and the first hadron, assumed to be the
leading hadron and $\sqrt{S}$ is the total center of mass energy. The parton
distribution functions are denoted by $f_{i/p} (x_1, \mu^2 )$ and are given by 
the CTEQ6 parameterization~\cite{CTEQ6}. $D^0_{h/i}(z_i,\mu^2)$ are the
unmodified fragmentation functions given by the KKP
parameterization~\cite{kkp}. All along this work, the scale $\mu^2$ is taken
to be the same for all the distribution and fragmentation functions and given
as the invariant mass squared of the final state hadrons, which we take as
pions. We will focus on the mid-rapidity region and thus, hereafter, set all
rapidities equal to zero. The momentum fractions $x_1, x_2, z_1, z_2$ are all
given in terms of $z_3$ and the transverse momenta of the produced hadrons. 
Furthermore, requiring that all momentum fractions are between $0$ and $1$ 
introduces the lower and upper limits on the $z_3$ integration. Further
details will be provided in Ref.~\cite{ajm2}.

\begin{figure}[t!] 
{\centering
{\epsfig{file=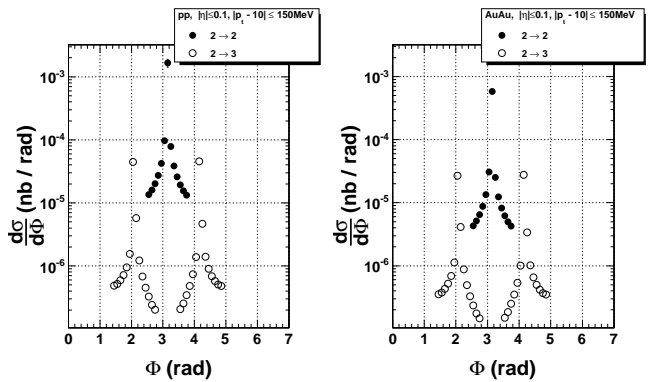, width=1\columnwidth}}
\par}
\caption{Comparison of the differential cross sections for two-to-two (full
  circles) and two-to-three (open circles) processes in p + p (left panel) and
  A + A (right panel)  collisions.} 
\label{fig1}
\end{figure}

We now use Eq.~(\ref{eq:pp}) to calculate the three hadron production cross
section in p + p for mid-rapidity at $\sqrt{S} =$ 200 GeV. The open circles on
the left panel of Fig.~\ref{fig1} show the differential cross section as a
function of the azimuthal angle $\phi$ for the particular case where each of
the three hadrons carry transverse momentum of $h_t=$ 10 GeV/c and therefore,
due to momentum conservation on partonic level, they are separated by an
angle of $\Delta\phi=2\pi/3$ radians. Only the angular distribution of 
the {\it away side} hadrons (with respect to the direction of one of the 
hadrons --the would be {\it leading} hadron-- which in the symmetric case
chosen here, can be any one of them) is shown.
 
In order to investigate the effects of a medium on three hadron production in
A + A collisions, we use the modified fragmentation function
proposed in~\cite{zoww} given by 
\be
   D_{h/i}(z_i,\mu^2)&=&(1-e^{-\langle \frac{L}{\lambda}\rangle}) 
   \left[ \frac{z_i^\prime}{z_i}D^0_{h/i}(z_i^\prime,\mu^2) +
   \langle \frac{L}{\lambda}\rangle \frac{z_g^\prime}{z_i} 
   \right. \nonumber\\
   &\times& \left.
   D^0_{h/g}(z_g^\prime,\mu^2)\right]
   + e^{-\langle\frac{L}{\lambda}\rangle} D^0_{h/i}(z_i,\mu^2),
\label{eq:mod_frag}
\ee
where $z_i^\prime= \frac{h_t}{(b_{ti}-\Delta E_i)}$ is the rescaled momentum
fraction of the leading parton with flavor $i$, $z_g^\prime=\langle
\frac{L}{\lambda} \rangle \frac{b_t}{\Delta E_i}$ 
is the rescaled momentum fraction of the radiated gluon, $\Delta E_i$ is the
average radiative parton energy loss and $\langle \frac{L}{\lambda}\rangle$ is
the average number of scatterings. The energy loss  $\Delta E_i$ is related
to the gluon density of the produced medium via
\be  
   \Delta E &=& \langle \frac{d E}{d L}\rangle_{1d} \, \int_{\tau_0}^{\infty} d
   \tau \frac{\tau - \tau_0}{\tau_0\, \rho_0}\, 
   \rho_g (\tau, \vec{b}_{\perp}, \vec{r}_t + \vec{n} \tau),
\label{eq:delE}
\ee
where $ \vec{b}_{\perp}$ (note that $b_t$ is used to denote the transverse
momentum of a parton) is the impact parameter of the collision. Since here we
consider the most central collisions, we set $ \vec{b}_{\perp} = 0$ in the
rest of the paper.  
$\vec{r}_t $ is the transverse plane location of the hard scattering where the
partons are produced and $\vec{n}$ is the direction in which the produced hard
parton travels in the medium. 
The average number of scatterings $\langle \frac{L}{\lambda}\rangle$, the one
dimensional energy loss $\langle \frac{d E}{d L}\rangle_{1d}$ and the gluon
density of the medium $\rho_g$ --which is related to the nuclear geometry of 
the produced medium-- are taken from Ref.~\cite{zoww} where we refer the
reader also for details on the chosen parameters.


The three produced partons will in general travel through the medium at
different angles with respect to the direction of $\vec{r}_t$ until they leave
the medium after which they hadronize. Therefore, the fragmentation
function for each produced parton will have a different dependence on the path
length traveled through the medium. Based on simple geometry one can show that
out of the three angles involved, only one is independent which we
take to be the angle that the direction of motion of particle 1 makes
with the position vector of the scattering center.
The normalized cross section is obtained from
Eq.~(\ref{eq:pp}) using the modified fragmentation functions given by
Eq.~(\ref{eq:mod_frag}), integrating over the location of the hard
scatterings and dividing by the nuclear overlap area. Due to 
the kinematics involved in mid-rapidity collisions, we expect the nuclear
modification of parton distribution functions to be small~\cite{nmc,e665} and
therefore ignore them.

The open circles on the right panel of Fig.~\ref{fig1} show the results of the
above analysis for the three hadron production differential
cross section as a function of azimuthal angle $\phi$ for mid-rapidity Au + Au
collisions at $\sqrt{S_{NN}}=200$ GeV. As in the p + p case, we also consider
that each of the three hadrons carry momentum with magnitude $h_t=$ 10
GeV/c. Since in our picture fragmentation occurs collinear to the direction of
the original parton, the final hadrons are separated by an angle
$\Delta\phi=2\pi/3$ radians. One can check that the ratio of the distributions
in the Au + Au to the p + p case is as a function of
$\phi$ approximately constant $\sim$ 0.7.  

In order to compare the three-hadron to the two-hadron production differential
cross section in the p + p and the A + A cases, we now compute the
latter as a function of the azimuthal angle $\phi$. To this end we use the LO
DIPHOX algorithm~\cite{DIPHOX}. We use CTEQ6 distribution
functions and in the p + p case, unmodified 
fragmentation functions given by the KPP parametrization. For the A + A case,
we use the modified fragmentation functions in Eq.~(\ref{eq:mod_frag})
integrating over the location of the hard scatterings and dividing by the
nuclear overlap area. To have a fair comparison to the already calculated
cross section for two-to-three processes, and similar to the experimental
situation, we take the leading 
and away side particles also having $h_t=10$ GeV/c. The full circles in
Fig.~\ref{fig1} show this cross section. On the left panel we show the results
for p + p collisions and on the right panel, the results of Au + Au, both for
$\sqrt{S_{NN}}=200$ GeV collisions at mid-rapidity. Notice that in both cases
the two-to-three hadron production cross 
section is suppressed with respect to the two-to-two result. However, and this
is the main result of our work, {\it this suppression is smaller in A + A
collisions than in p + p collisions}. Dividing this ratio in A + A to that in
p + p, we get as a function of $\phi$ approximately a constant 
$\sim$ 2.26.  

Since in the present analysis, the sole ingredient that distinguishes between 
the p + p and the A + A cases is the energy loss of partons that hadronize
collinearly, then the only reason for the cross sections in two-to-three
processes to be less suppressed in A + A than in p + p, when compared to the
two-to-two processes, is the different geometry for the trajectories of three
as opposed to two particles in the final state.


\begin{figure}[t!] 
{\centering
{\epsfig{file=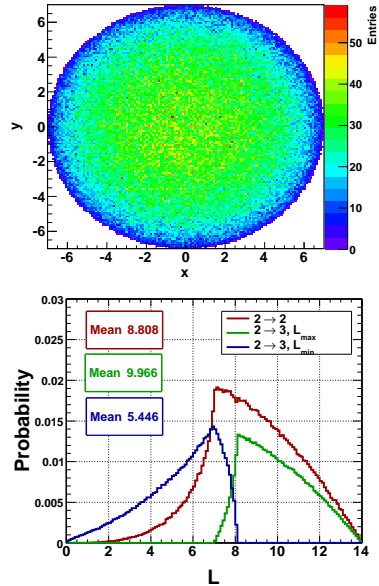, width=0.6\columnwidth}}
\par}
\caption{(Color online) Distribution of scattering centers (upper panel) and
 path lengths for two-to-two and two-to-three processes (lower panel) for the
 {\it away side} partons, obtained by discarding the shortest of the path
 lengths in each case. When three partons are present in the final
 state, the away side parton with the average large (short) path length 
 has a greater (smaller) path length than the away side parton in the case with
 two partons in the final state.}   
\label{fig2}
\end{figure}

To test this idea, we compute the distribution of path lengths with two
and three hadrons in the final state. We take a nuclear overlap area
with a distribution of scattering centers denser in the middle and decreasing
toward the edge. In each case we disregard the shortest path length (the one
that would correspond to the trigger particle) and compute the distribution of
the other path lengths (the ones corresponding to the away side
particles). This is shown in Fig.~\ref{fig2}. The upper panel shows the
distribution of scattering centers and the lower one the distribution of path
lengths. Notice that indeed, when there are three hadrons  in the final state,
the large (short) path length in the away side is greater (smaller) than the
case of the away side particle when there 
are two hadrons in the final state. This means that for the case of 
three particles in the final state, even if one of the non-leading particles
gets absorbed by the medium (the one with the largest path length), the
remaining particle has a large probability of punching through than in the
case when one has two particles in the final state. We interpret this as
signaling that in two-to-three processes, there is a large probability to have
one of the two away side particles being absorbed an the other randomly
getting out, producing on the average, a double hump.


In conclusion, we have presented a LO calculation in p + p collisions for the
cross section of two-to-three processes and compared to the corresponding
cross sections for two-to-two processes, using the same approximations. We
have also computed the same processes in A + A collisions taking into account
a detailed Glauber approach, similar to the one employed in
Ref.~\cite{Dainese}. The results show that the cross section in
two-to-three processes in A + A is less suppressed --with respect to two-to-two
processes-- than in the case of p + p. We attribute this result to a purely
geometrical effect associated with the differences in the path
lengths of three as opposed to two particles in the final state in the case of
A + A collisions. These results raise several interesting possibilities that
need to be addressed, among them we can point out the following: 1)
Two-to-three processes are naturally expected and the strenght of their signal
is not beyond the capabilities of current experiments. 2) 
Considering that two-to-three processes exist in p + p collisions, as
suggested also in Ref.~\cite{Ayala}, our work shows that their
observation in A + A should be enhanced with respect to the strongly
energy-loss suppressed two-to-two processes and thus, that this effect may have
a bearing on the shape of the away side for different kinematical cuts in
azimuthal correlations in A + A. A detailed study of this effect is currently
under way and will be reported elsewhere.

\section*{Acknowledgments} 

A.A. and J.J-M. thank CBPF for their hospitality and
financial support during a visit when this work was initiated. Support has
been received in part by DGAPA under PAPIIT grant 
IN116008. J.J-M. thanks A. Dumitru for discussions and is supported by
The City University of New York through a PSC-CUNY research grant and by the
Office of Nuclear Physics, US-DOE Grant DE-FG02-09ER41620. J.M. is
supported by FAPERJ, under Proj. E-26/110.266/2009 CNPq, the Brazilian
Council for Science and Technology. M.E.T.-Y. thanks the {\it Programa de
Intercambio UNAM-UNISON} for support.


\begin{thebibliography}{55}

\bibitem{eloss}
 M.~Gyulassy and X.~N.~Wang,
  Nucl.\ Phys.\  B {\bf 420}, 583 (1994)
  [arXiv:nucl-th/9306003]; 
 R.~Baier, Y.~L.~Dokshitzer, A.~H.~Mueller, S.~Peigne and D.~Schiff,
  Nucl.\ Phys.\  B {\bf 484} (1997) 265
  [arXiv:hep-ph/9608322]; 
M.~Gyulassy, P.~Levai and I.~Vitev,
  Nucl.\ Phys.\  B {\bf 594}, 371 (2001)
  [arXiv:nucl-th/0006010]; 
 U.~A.~Wiedemann,
  Nucl.\ Phys.\  B {\bf 588}, 303 (2000)
  [arXiv:hep-ph/0005129]; 
 S.~Wicks, W.~Horowitz, M.~Djordjevic and M.~Gyulassy,
  Nucl.\ Phys.\  A {\bf 784}, 426 (2007)
  [arXiv:nucl-th/0512076].

\bibitem{urs}
For a recent review, see
U.~A.~Wiedemann,
  arXiv:0908.2306 [hep-ph].

\bibitem{rhic_exp}
I.~Arsene {\it et al.}  [BRAHMS Collaboration],
  Nucl.\ Phys.\  A {\bf 757}, 1 (2005); 
 B.~B.~Back {\it et al.},
  Nucl.\ Phys.\  A {\bf 757}, 28 (2005); 
 K.~Adcox {\it et al.}  [PHENIX Collaboration],
  Nucl.\ Phys.\  A {\bf 757}, 184 (2005); 
 J.~Adams {\it et al.}  [STAR Collaboration],
  Nucl.\ Phys.\  A {\bf 757}, 102 (2005).

\bibitem{gm}
 M.~Gyulassy and L.~McLerran,
  Nucl.\ Phys.\  A {\bf 750}, 30 (2005)
  [arXiv:nucl-th/0405013].

\bibitem{onehad}
  S.~Jeon, J.~Jalilian-Marian and I.~Sarcevic,
  Phys.\ Lett.\  B {\bf 562}, 45 (2003), 
  Nucl.\ Phys.\  A {\bf 723}, 467 (2003).

\bibitem{PHENIX}
A. Adare et. al, (PHENIX Collaboration), Phys. Rev. C {\bf 78}, 014901 (2008).

\bibitem{STAR}
F. Wang, et al. (STAR Collaboration), Nucl. Phys. A {\bf 774}, 129 (2006).

\bibitem{Casalderrey}
J. Casalderrey-Solana, E.V. Shuryak, D. Teaney, J. Phys. Conf. Ser. {\bf 27},
22 (2005). 

\bibitem{miklos}
G.~Torrieri, B.~Betz, J.~Noronha and M.~Gyulassy,
Acta Phys.\ Polon.\  B {\bf 39}, 3281 (2008). 

\bibitem{Majmunder}
A. Majumder, X.-N. Wang, Phys. Rev. C {\bf 73}, 051901(R) (2006).

\bibitem{Chiu}
C. B. Chiu, R. C. Hwa, Phys. Rev. C {\bf 74}, 064909 (2006).

\bibitem{Chaudhuri}
A. K. Chaudhuri, Phys. Rev. C {\bf 77}, 027901 (2008).

\bibitem{Takahashi}
J. Takahashi et al., {\it Topology studies of hydrodynamics using two particle
correlation analysis}, arXiv:0902.4870.

\bibitem{Ayala}
A. Ayala, E. Cuautle, I. Dom\'{\i}nguez, A. Ort\'{\i}z and G. Pai\'c,
Eur. Phys. J. C {\bf 62}, 535 (2009).

\bibitem{lo}
  R.~K.~Ellis and J.~C.~Sexton,
  Nucl.\ Phys.\  B {\bf 269}, 445 (1986); 
  F.~A.~Berends, R.~Kleiss, P.~De Causmaecker, R.~Gastmans and T.~T.~Wu,
  Phys.\ Lett.\  B {\bf 103}, 124 (1981).

\bibitem{zoww}
  H.~Zhang, J.~F.~Owens, E.~Wang and X.~N.~Wang,
  Phys.\ Rev.\ Lett.\  {\bf 98}, 212301 (2007).


\bibitem{CTEQ6}
J. Pumplin, D.R. Stump, J.Huston, H.L. Lai, P. Nadolsky and W.K. Tung, JHEP
{\bf 0207}, 012 (2002).


\bibitem{kkp}
 B.~A.~Kniehl, G.~Kramer and B.~Potter,
  Nucl.\ Phys.\  B {\bf 582}, 514 (2000)
  [arXiv:hep-ph/0010289].

\bibitem{ajm2}
A.~Ayala, J.~Jalilian-Marian, J.~Magnin, A. Ort\1z, G. Pai\'c and
M. E. Tejeda-Yeomans, in progress.

\bibitem{nmc}
M.~Arneodo {\it et al.}  [New Muon Collaboration],
  Nucl.\ Phys.\  B {\bf 481}, 23 (1996), 
 Nucl.\ Phys.\  B {\bf 441}, 12 (1995), 
 Nucl.\ Phys.\  B {\bf 481}, 3 (1996).

\bibitem{e665}
M.~R.~Adams {\it et al.}  [E665 Collaboration],
  Phys.\ Rev.\  D {\bf 54}, 3006 (1996). 

\bibitem{DIPHOX}
T. Binoth, J. Ph. Guillet, E. Pilon and M. Werlen, Eur. Phys. J. C24, 245
(2002).

\bibitem{Dainese}
A. Dainese, C. Loizides and G. Pai\'c, Eur. Phys. J. C {\bf 38}, 461 (2005).

\end{thebibliography}
\end{document}